\documentclass[twocolumn,prd,superscriptaddress,nofootinbib]{revtex4-1}
\usepackage[dvipsnames]{xcolor} 
\usepackage{amsmath} 
\usepackage{booktabs}
\usepackage[utf8]{inputenc}
\usepackage{graphicx} 
\usepackage[colorlinks=true,citecolor=blue,urlcolor=blue]{hyperref}
\usepackage[normalem]{ulem}
\usepackage{xspace}
\usepackage[caption=false]{subfig}
\usepackage{orcidlink}
\usepackage{multirow}

\newcommand{\dcc}{LIGO-P}

\def\prd{{\it Physical Review D}\xspace}

\begin{document}

\title{A neural network-based gravitational wave interpolant with applications to low-latency analyses}
\author{Ryan Magee \orcidlink{0000-0001-9769-531X}}
\email{rmmagee@caltech.edu}
\affiliation{LIGO, California Institute of Technology, Pasadena, CA 91125, USA}

\author{Richard George \orcidlink{0000-0002-7797-7683}}
\affiliation{Center for Gravitational Physics, University of Texas, Austin, TX 78712, USA}

\author{Alvin Li \orcidlink{0000-0001-6728-6523}}
\affiliation{LIGO, California Institute of Technology, Pasadena, CA 91125, USA}

\author{Ritwik Sharma \orcidlink{0000-0003-1858-473X}}
\affiliation{Department of Physics, Deshbandhu College, University of Delhi, New Delhi, India}
\affiliation{STAR Lab, Indian Institute of Technology Bombay, Powai 400076, India}
\begin{abstract}

Matched-filter based gravitational-wave search pipelines identify candidate
events within seconds of their arrival on Earth, offering a chance to guide
electromagnetic follow-up and observe multi-messenger events.  Understanding
the detectors' response to an astrophysical transient across the searched signal
manifold is paramount to inferring the parameters of the progenitor and
deciding which candidates warrant telescope time. 
We describe a framework that uses artificial
neural networks to interpolate gravitational waves and, equivalently, the signal-to-noise ratio
(SNR) across sufficiently local patches of the signal manifold. Our
machine-learning based model generates a single waveform in 6 milliseconds on a
CPU and 0.4 milliseconds on a GPU. When using a GPU
to generate batches of waveforms simultaneously, we find that we can produce
$10^4$ waveforms in $\lesssim 1$ ms. This is achieved while remaining faithful, on
average, to 1 part in $10^4$ (1 part in $10^5$) for binary black hole (binary
neutron star) waveforms. The model we present is designed to directly utilize
intermediate detection pipeline outputs in the hopes of facilitating a
better real-time understanding of gravitational-wave candidates. 

\end{abstract}

\maketitle

\section{Introduction}\label{sec:intro}

Gravitational-waves (GW) observed by Advanced
LIGO~\cite{LIGOScientific:2014pky} and Advanced Virgo~\cite{VIRGO:2014yos} have
provided an unprecedented view into our Universe. The candidates identified by
the LIGO-Virgo-Kagra collaboration (LVK)~\cite{LIGOScientific:2021djp} and external
groups~\cite{Nitz:2018imz,Magee:2019vmb,Venumadhav:2019lyq,Zackay:2019btq,Nitz:2020oeq,Nitz:2021uxj,Nitz:2021zwj,Olsen:2022pin}
provide an increasingly complete census of the ultracompact binaries in our
universe~\cite{KAGRA:2021duu}. Much of this science has been made possible by
the low-latency search
pipelines~\cite{Tsukada:2023edh,Nitz:2018rgo,Aubin:2020goo,Chu:2020pjv,Drago:2020kic}
that rapidly identifies GW candidates and the alert infrastructure~\cite{Magee:2021xdx,Chaudhary:2023vec}
that subsequently localizes and classifies them.

Binary neutron stars (BNSs) and neutron
star -- black hole binaries (NSBHs) remain particularly sought after since they
are known~\cite{LIGOScientific:2017vwq,LIGOScientific:2017ync} (or, in the case
of NSBHs, expected~\cite{PhysRevD.98.081501}) sources of electromagnetic
radiation.  It is therefore crucial to precisely understand the nature of compact
binaries in near-real-time to determine which warrant electromagnetic followup.
This is increasingly important as highly anticipated, next generation
facilities like the Rubin
Observatory~\cite{Ivezic:2008fe} begin observations in the coming years. At
present, our low-latency estimates of localization~\cite{Singer:2015ema},
classification~\cite{2015PhRvD..91b3005F,Kapadia:2019uut,Andres:2021vew,Villa-Ortega:2022qdo},
and source properties
~\cite{PhysRevD.98.081501,PhysRevD.104.083003} are derived from the highest
signal-to-noise ratio (SNR) event among the maximum likelihood candidates identified by LVK
detection pipelines. Though this results in precise
localizations~\cite{Singer:2014qca}, it is well understood that full parameter
estimation is systematically more accurate~\cite{Chaudhary:2023vec}.
Unfortunately, even in their low-latency configuration, parameter estimation pipelines can take
$\mathcal{O}(\mathrm{hours})$ to produce stable results, which may be too late
to facilitate the capture of rapidly fading transients~\cite{Cowperthwaite:2017dyu}. 

In this work, we examine neural networks as a way to interpolate the search
response, in the hopes of developing a better real-time understanding of the
nature of the candidate and enabling rapid parameter estimation in
the future.  Recently, machine-learning algorithms have emerged as a promising
way
to navigate GW science and other areas of data-driven astrophysics.  This has
largely been made possible by advances to graphical processing units (GPUs) and
the ease of software that utilize them~\cite{cuda,NEURIPS2019_9015,xla}. In GW
astrophysics alone, machine-learning algorithms have already been applied to
compact binary
detection~\cite{George:2017pmj,Gabbard:2017lja,Schafer:2022dxv,Mishra:2022ott,Krastev:2019koe,Schafer:2020kor,Schafer:2022dxv,Nousi:2022dwh,Marx:2024wjt},
early warning
alerts~\cite{Wei:2020sfz,Baltus:2021nme,Yu:2021vvm,Baltus:2022pep,Alfaidi:2024ioo}, supernovae
identification~\cite{Astone:2018uge,Antelis:2021qak,LopezPortilla:2020odz},
parameter
estimation~\cite{Gabbard:2019rde,Dax:2021tsq,Chatterjee:2022ggk,Chatterjee:2022dik,Dax:2022pxd,Wong:2023lgb},
noise removal and
characterization~\cite{Vajente:2019ycy,Essick:2020qpo,Saleem:2023hcm}, and waveform interpolation~\cite{Schmidt:2020yuu,Chua:2018woh,Khan:2020fso,Tissino:2022thn,Thomas:2022rmc}.

Previous waveform interpolation applications have primarily focused on aligned-spin
binary black holes (BBH)~\cite{Chua:2018woh,Khan:2020fso,Schmidt:2020yuu}, though some have
considered non-aligned systems as well~\cite{Thomas:2022rmc}. Each of these
works first converts the waveforms to a new basis, which is achieved via
methods such as principal component analysis and Gram-Schmidt
orthonormalization, before interpolating the coefficients needed to reconstruct
the waveforms. In this work, we describe a similar
technique. We employ neural networks trained via
supervised learning to the interpolation of the GW emission and
SNR of arbitrary aligned-spin compact
binaries using singular value decomposition (SVD). This work is inspired by
previous studies that used
grid-based techniques to interpolate the SVD of nonspinning BBH
signals~\cite{Cannon:2011rj}, as well as mesh-free approaches applied to the
SVD of aligned-spin BNS emission~\cite{Pathak:2022ktt}.
We build on previous SVD interpolation
efforts and additionally consider the impacts of component spin,
interpolate waveforms of arbitrary length, and provide a method that is
immediately applicable to existing low-latency analyses and requires minimal
fine tuning.  The interpolant we describe provides a way to rapidly
calculate the SNR of aligned-spin compact binaries, which may be able to
accelerate low-latency data enrichment tasks.

This paper is organized as follows. First, we describe the matched-filtering
algorithms and neural network architecture in Section~\ref{sec:methods}.
Second, we apply this network to BBH and BNS 
systems. We describe the accuracy of the interpolant on these systems and present a timing study in Section~\ref{sec:study}. Finally, we conclude with a discussion of
future applications in low-latency environments in Section~\ref{sec:conclusion}.

\section{Methods}
\label{sec:methods}
\subsection{Background and motivation}

Modern matched-filter based GW pipelines simultaneously search for GWs from
BNS, BBH, and NSBH binaries. The searches precompute
the expected gravitational wave emission, also known as a \emph{template}, 
for $\mathcal{O}(10^6)$ binaries representative of the space. This collection
of templates, or \emph{template bank}, is designed to retain $97\%-99\%$ of the
SNR. While the overall parameter space is
vast, pipelines often divide their search across a number of
small, partially-overlapping \emph{sub-banks} in the signal manifold that contain
templates with morphologies~\cite{Messick:2016aqy,Roulet:2019hzy,Sakon:2022ibh}
that respond similarly to astrophysical sources and detector noise.

The similarities between the waveforms in a sub-bank can be exploited by using
\emph{singular value decomposition} (SVD)~\cite{Cannon:2010qh,Roulet:2019hzy}
to reduce the
filtering cost of the search. The SVD ensures that we can express any of the
$N$\footnote{Performing an SVD on large matrices is known to be computationally
expensive. Within the GstLAL framework, searches set $ N \leq 1024 $ to reduce cost.} waveforms within
our sub-bank as a sum of some orthogonal basis vectors, $u_\mu(t)$, and
complex-valued reconstruction coefficients, $a(\lambda)_\mu$:
\begin{equation} h(\lambda, t) = \sum_{\mu=0}^{M} a(\lambda)_\mu u_\mu(t) \end{equation}
where $\lambda= (m_1, m_2, s_{1z}, s_{2z})$ represents the set of intrinsic parameters of the system, $h(\lambda,t)$ is the
whitened gravitational-wave strain for parameters $\lambda$ at time $t$, $M$ represents the number of bases identified by the SVD, and $M \leq N$.
We neglect the impact of the location and orientation of the binary on
the GW emission, which can be applied via a complex rotation for
aligned-spin systems. Further approximations can be applied to ensure that $M
<< N$, which greatly
reduces the number of filters a search must correlate with the data. We refer
the reader to~\cite{Cannon:2010qh} for a complete description of this method applied to compact binary searches.

A further speed up can be achieved by first
segmenting the waveforms into distinct time-slices and downsampling within each
slice to the Nyquist rate associated with the highest frequency component:
\begin{equation} h(\lambda,k) = \sum_{s=0}^{S-1}
\begin{cases} h(\lambda,k\frac{f}{f^s})^{s} & t^s \leq k/f^s < t^{s+1} \\
0 & \rm{else}
\end{cases}
\end{equation}
where $h(\lambda,t)^s$ is the
whitened gravitational-wave strain at time $t$ in time slice $s$, $f^s$ is
the sample-rate in the corresponding time slice, and $S$ is the total number of time slices.

Each time slice is independent of the next, and a SVD
can be performed on each individually to identify orthonormal bases for each
segment:
\begin{equation} h(\lambda, t)^{s} = \sum_{\mu=0}^{M-1} a_\mu^{s}(\lambda) u_\mu^{s}(t) \end{equation}
Here, $a_\mu^s(\lambda)$ denotes the complex\footnote{The complex valued coefficient encodes the two polarizations
of a GW under general relativity.} valued reconstruction coefficient
associated with basis vector $\mu$ and time slice $s$, $u_\mu^s(t)$ is the
basis vector for time slice $s$, and $M$ is the number of basis vectors
associated with time slice $s$. This process is known as the Low Latency Online
Inspiral Detection (LLOID) algorithm, and it is currently employed within the
GstLAL-based detection pipeline. An comprehensive description of the LLOID
algorithm can be found in~\cite{Cannon:2011vi}. A similar, frequency domain
procedure is used by MBTA~\cite{Adams:2015ulm}. 

Previous studies have provided numerical evidence that bases
produced by SVD are complete over
the signal manifold from which they were derived~\cite{Cannon:2011xk}.
Our goal in this work is exploit that completeness and to
use machine learning to faithfully interpolate the
reconstruction coefficients $a_\mu^{s}(\lambda)$ for any $\lambda$
contained within the corresponding sub-bank. 
Note that under the LLOID framework, the SNR time series can be written as:
\begin{equation}\label{eq:snr} \rho(\lambda, t)^s = \langle h(\lambda, t)^{s} | d(t) \rangle =
\sum_{\mu=0}^{M-1} a_\mu^{s}(\lambda) \langle u_\mu^{s}(t) | d \rangle
\end{equation}
where, $\langle \cdot | \cdot \rangle$ denotes the noise weighted inner product
\begin{equation}
\langle a(t) | b(t) \rangle \equiv 4 \int_0^{\infty}
\mathrm{d}f\frac{a^*(f)b(f)}{S_n(f)} \end{equation} for the one-sided power
spectral density (PSD), $S_n(f)$. The final term in equation~\ref{eq:snr}, $\langle
u_\mu^{s}(t) | d (t)\rangle$, is the orthogonal SNR measured directly by the
pipeline. The SNR corresponding to the physical template, $\rho$, is
reconstructed via matrix multiplication.  Interpolating the reconstruction coefficients $a_\mu^s(\lambda)$ is therefore equivalent to producing arbitrary waveforms \emph{and} arbitrary SNR timeseries since both utilize the same coefficients. Interpolating the SNRs across the signal
manifold exposes more information than would otherwise be available from
the small, discrete space sampled by the search, and will enrich our
understanding of the properties of low-latency GW candidates.

\subsection{Machine learning architecture}
\label{sec:ml}

We use a feed-forward neural network as a form of supervised learning to derive
the relationship between our four inputs, $\lambda$, and our two real-valued outputs
per basis vector, $\operatorname{Re}\, a_\mu^{s}(\lambda)$ and
$\operatorname{Im}\, a_\mu^{s}(\lambda)$. Such networks consist of three types
of layers (input, hidden, and output), where each layer contains nodes, or
\emph{neurons}, that are fully connected to adjacent layers of arbitrary size.
The input to any layer can be represented by a vector $\vec{x}$ with
dimensionality equal to the number of neurons. The interaction between a layer with input
size $N$ and output size $M$ can be completely characterized by a weight
matrix, $W$, with dimension $N \times M$, a bias vector, $\vec{b}$, and an
activation function, $f(x)$, which is typically non-linear and is applied to
each output neuron individually. The weight matrix describes the strength of the
connection between any two neurons in adjacent
layers, while the bias term provides added
flexibility to the model by providing an offset.  The output for a given neuron
is then
\begin{equation}
\vec{y} = f(W^T\thinspace\vec{x}  + \vec{b})
\end{equation}
where $\vec{y}$ has dimension $M$.

The
weights and biases that describe a network are not set initially, but learned
via supervised training. This learning occurs by minimizing a loss function
that compares the target training data to the network output; the most commonly
used loss function for regression-based tasks such as the one described here is the mean-square error,
\begin{equation}
\mathrm{MSE} = \frac{1}{n}\sum_{i = 0}^{n} (y_i - \hat{y}_i)^2
\end{equation}
where $y_i$ and $\hat{y}_i$ are the expected values from the
training data and the network outputs, respectively, and $n$ denotes the size of the dataset.

\begin{figure}
\includegraphics[width=\columnwidth]{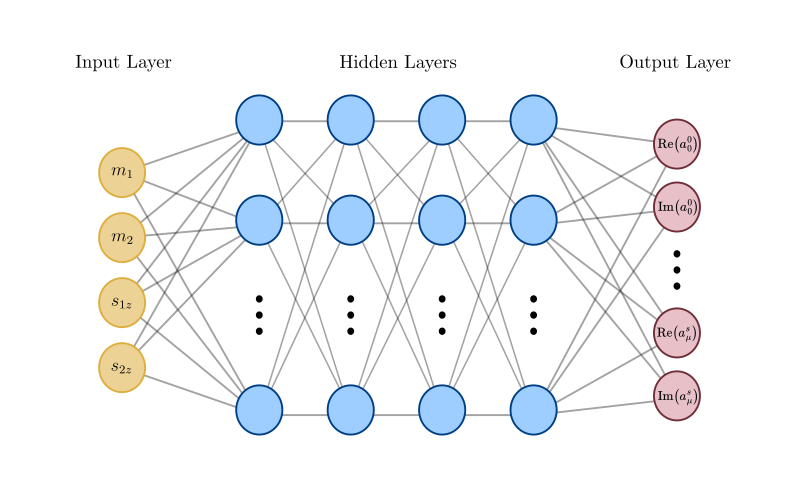}
\caption{A schematic of the neural network architecture used in this work. The
network takes the intrinsic parameters $\lambda = (m_1, m_2, s_{1z},
s_{2z})$ as input and returns the reconstruction coefficients
$a_\mu^s(\lambda)$ as output. Our network has two outputs per basis vector,
representing the two polarizations of the GW. The total number of outputs
depends on the number of time slices and SVD bases needed for a particular sub-bank.}
\label{fig:architecture}
\end{figure}

The appropriate choice for the number of layers, neurons per layer, and
activation function can vary depending on the task, though there are a number of
key features that are typically desired. Continuous differentiability, monotonicity,
and fixed output ranges are generally preferred since they facilitate network
backpropagation, gradient descent, and learning stability, and can ameliorate the
vanishing or exploding gradient problem~\cite{Goodfellow-et-al-2016}. In this
work,
we use the network to perform regression and estimate a set of weights that
describes the contribution for a number of basis vectors. We use a neural
network consisting of a 4-node input layer corresponding to $\lambda = (m_1, m_2, s_{1z},
s_{2z})$, 4 hidden layers of 512 neurons, and an output layer whose size
depends on the total number of basis vectors contained in a specific sub-bank
(Fig.~\ref{fig:architecture}). For our data, each time slice is independent of
the next, so we have the freedom to fit each segment of the waveform
separately.  Similarly, we could train each basis vector independently.
Instead, we elect to fit all time slices and bases at once to reduce the total
number of models needed. We use Rectified Linear Units (ReLUs) as the
activation function for our hidden layers, defined as
\begin{equation}
f(x) = \mathrm{max}(0,x) \,.
\end{equation}
We use the mean-square error as our
loss function, and the \texttt{Adam} optimizer~\cite{Kingma:2014vow} to
minimize the loss function. We choose to use $\lambda$ as
the input to our network to highlight the generality of this method, but we note that transforming coordinates to
the parameters most important for the phasing of the
signal~\cite{Cutler:1994ys,Morisaki:2020oqk} could lead to more accurate or faster converging results. We do not exclude the possibility that another set of coordinates may be necessary in regions of the parameter space that exhibit significant degeneracies.

\section{Study}
\label{sec:study}
\subsection{Fidelity}
We apply the above formalism to BNS and BBH systems.
We generate small BNS and BBH banks with minimum
matches~\cite{Owen:1998dk} of $0.99$ using a stochastic placement
algorithm~\cite{Harry:2009ea,Privitera:2013xza}. The banks are not
intended to be representative of a full GW search; rather, they are
constructed in a constrained section of the parameter space to be
illustrative of sub-bank grouping in current low-latency
analyses. We refer the reader to Figures 2 and 4
in~\cite{Sachdev:2019vvd} and~\cite{Sakon:2022ibh}, respectively, for grouping
visualizations. The BBH (BNS) bank is designed to recover compact binaries
with $ 20 M_\odot < m_2 \leq
m_1 < 40 M_\odot$
($ 1.40 M_\odot < m_2 \leq m_1 < 1.42 M_\odot $) and $ 0.5 <
s_{iz} < 0.75 $ ($ 0.00 < s_{iz} < 0.01$), where $m_1, m_2$ are the
component masses of the binary and $s_{1z}, s_{2z}$ are the components of the
spin angular momentum aligned with the orbital angular momentum.  The resulting
bank contains 193 (343) waveforms. We verify the coverage of the banks by
drawing $10^4$ random samples from within the bounds and computing the mismatches,
\begin{equation}
\mathcal{M}(h_i, h_j) \equiv 1 - \frac{\langle h_i | h_j \rangle}{\langle h_i | h_i \rangle \langle h_j | h_j \rangle}
\end{equation}
between each sample and every template within the respective bank. In each case,
no simulation has a mismatch $\geq .011$ with some template in the bank. The
median mismatch across all simulations is 0.002 (0.003) for the BBH (BNS) bank.

We model the gravitational wave emission in each bank using the \texttt{SEOBNRv4\_ROM}
(\texttt{TaylorF2}) waveform approximants~\cite{Bohe:2016gbl,Blanchet:1995ez}, which are presently used to model the BBH (BNS) region of the GstLAL template bank in Advanced LIGO, Advanced Virgo, and KAGRA's fourth observing run~\cite{Sakon:2022ibh}. The emission is modeled 
from a starting frequency of 20 Hz (32 Hz) to capture the entirety of the BBH signals and the last minute of the BNS
signals in the detectors' sensitive bands. For each bank, we use the LLOID algorithm to first
segment the waveforms in time, and then to find a set of orthogonal basis
vectors for each time-slice.  This yields 2 (7) time slices with a total of 62
(80) basis vectors distributed across them. 

Ultimately, our network aims to predict $a_\mu^s(\lambda)$ for each basis,
$\mu$, and time slice, $s$. To construct data to train our machine-learning
model, we uniformly draw $10^5$ samples from the 4-dimensional hyper-rectangle
defined by the boundaries of each template bank.  Similar
work~\cite{Wong:2020wvd} has utilized Latin Hypercube
sampling~\cite{OLSSON200347} to provide samples that are representative
of variability within the hypercube. Since our interpolant is meant to act in a
highly parallel fashion and over local areas of the signal manifold where there
is not substantial metric variability, we find that
random sampling suffices. We segment and whiten each waveform using the publicly
available \texttt{aligoO4\_low.txt} estimate of Advanced LIGO's power spectral density for the fourth
observing run\footnote{\href{https://dcc.ligo.org/LIGO-T2000012/public}{https://dcc.ligo.org/LIGO-T2000012/public}.}.
We rotate the waveforms so that the phase at the time of peak amplitude is zero\footnote{This convention is chosen so that our interpolated waveforms all have phase zero and can be trivially rotated to arbitrary phase.},
and then project each waveform segment onto the corresponding basis vector
obtained from the SVD to obtain the complex valued reconstruction coefficients
associated with each time slice and basis:
\begin{equation} a_\mu^{s}(\lambda) = h^{s}(\lambda,t) \cdot u_\mu^{s}(t) \end{equation}
We store the reconstruction coefficients as a pair of real-valued numbers, $\operatorname{Re}\,
a_\mu^{s}(\lambda)$ and $\operatorname{Im}\, a_\mu^{s}(\lambda)$. The number of coefficients
needed to reconstruct the original waveform scales with the
numbers of bases and the
number of time slices. Though in general this quantity changes
across the parameter space, it remains constant within a single
decomposed bank. 

\begin{figure}
\includegraphics[width=\linewidth]{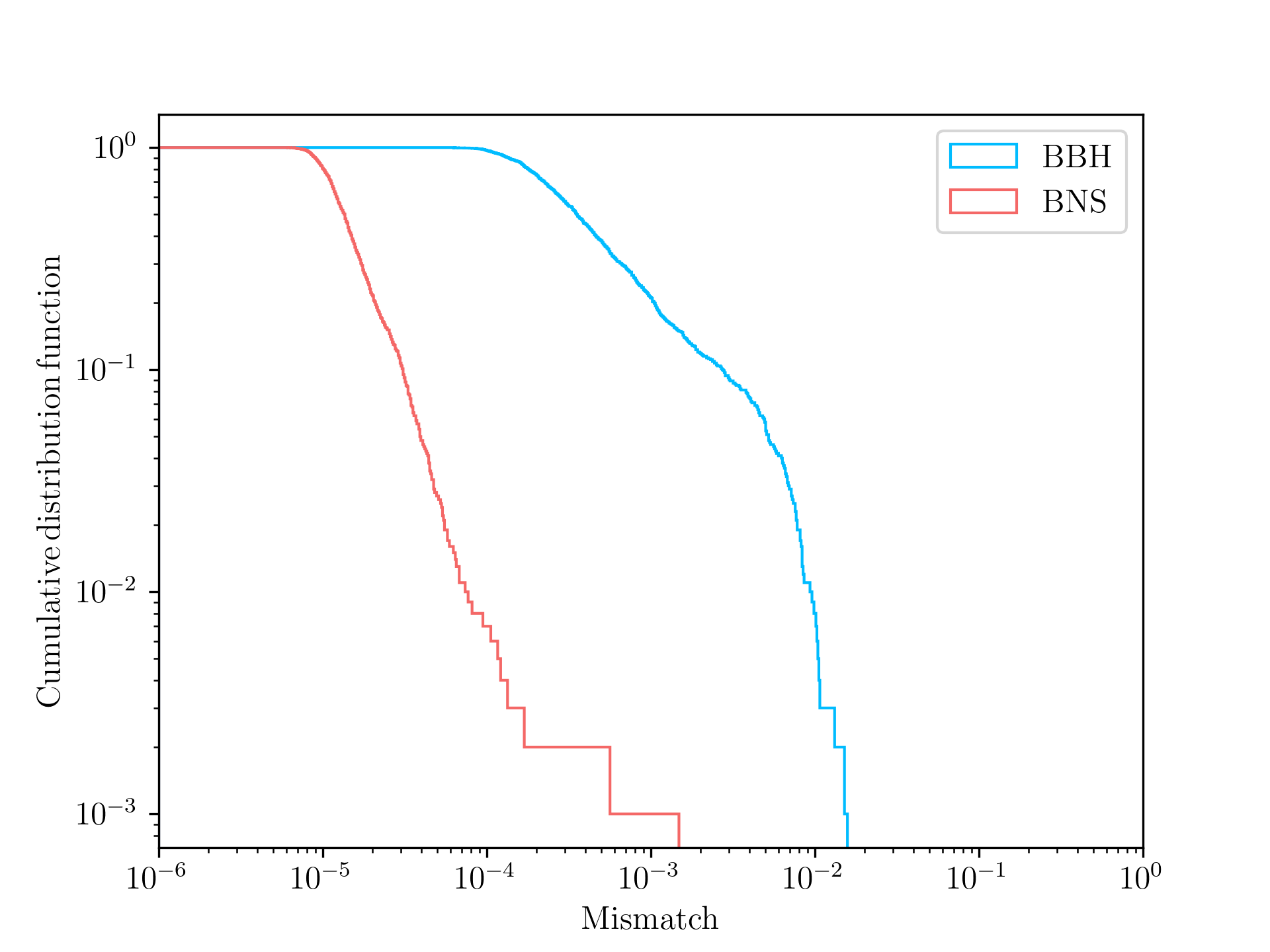}
\caption{\label{fig:mismatch} Reconstruction accuracy of the neural-net
interpolated waveform for BBH and
BNS systems. All interpolated waveforms have a mismatch no worse than
one part in $10^{2} (10^{3})$ and, on average, 1 part in $10^4 (10^5)$ for BBH (BNS), which is at the level of waveform systematics.}
\end{figure}

\begin{figure*}
\centering
\subfloat[BBH]{%
	\includegraphics[width=0.48\linewidth]{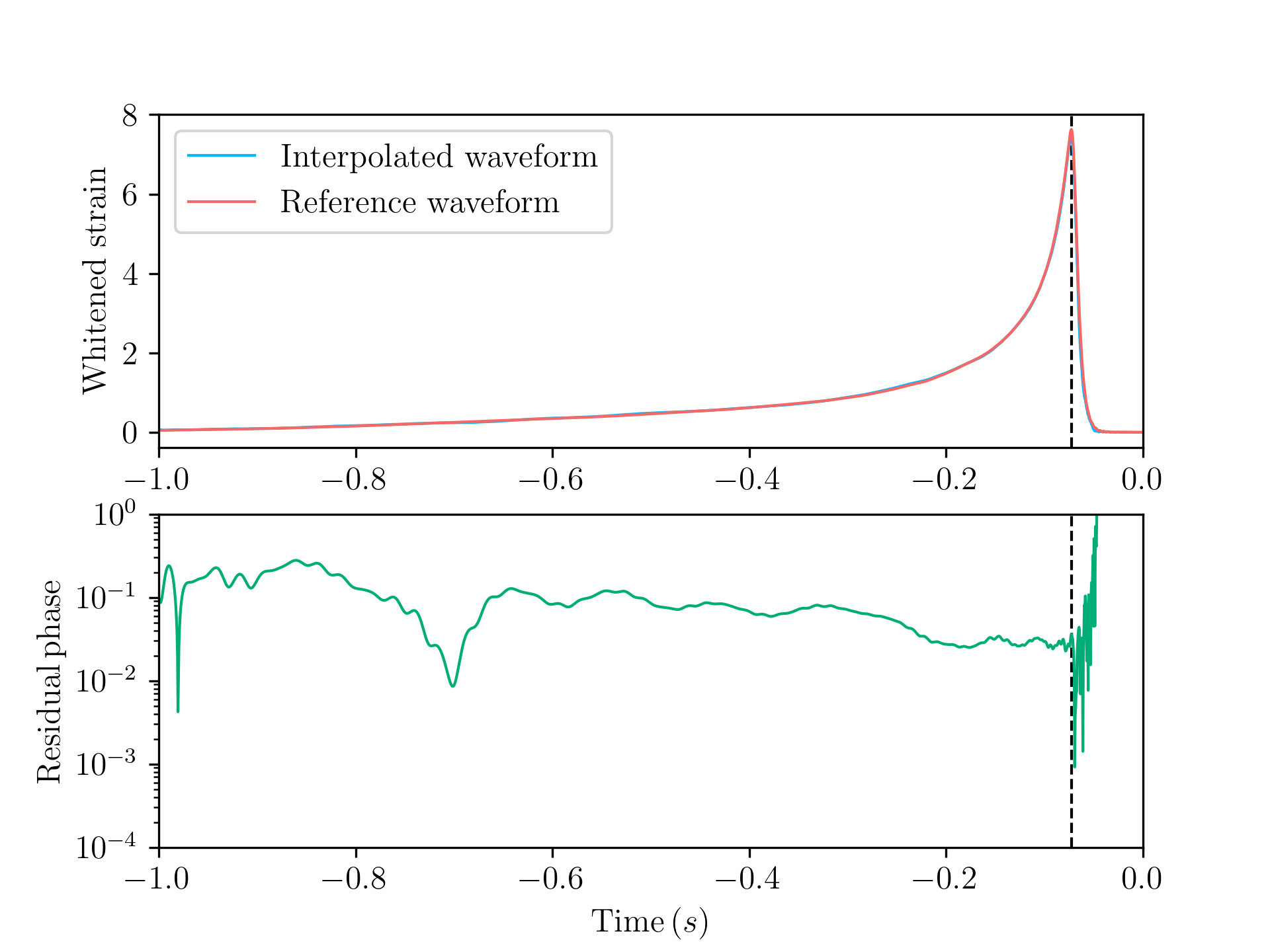}
	\label{fig:bbh_residual}
}
\subfloat[BNS]{%
	\includegraphics[width=0.48\linewidth]{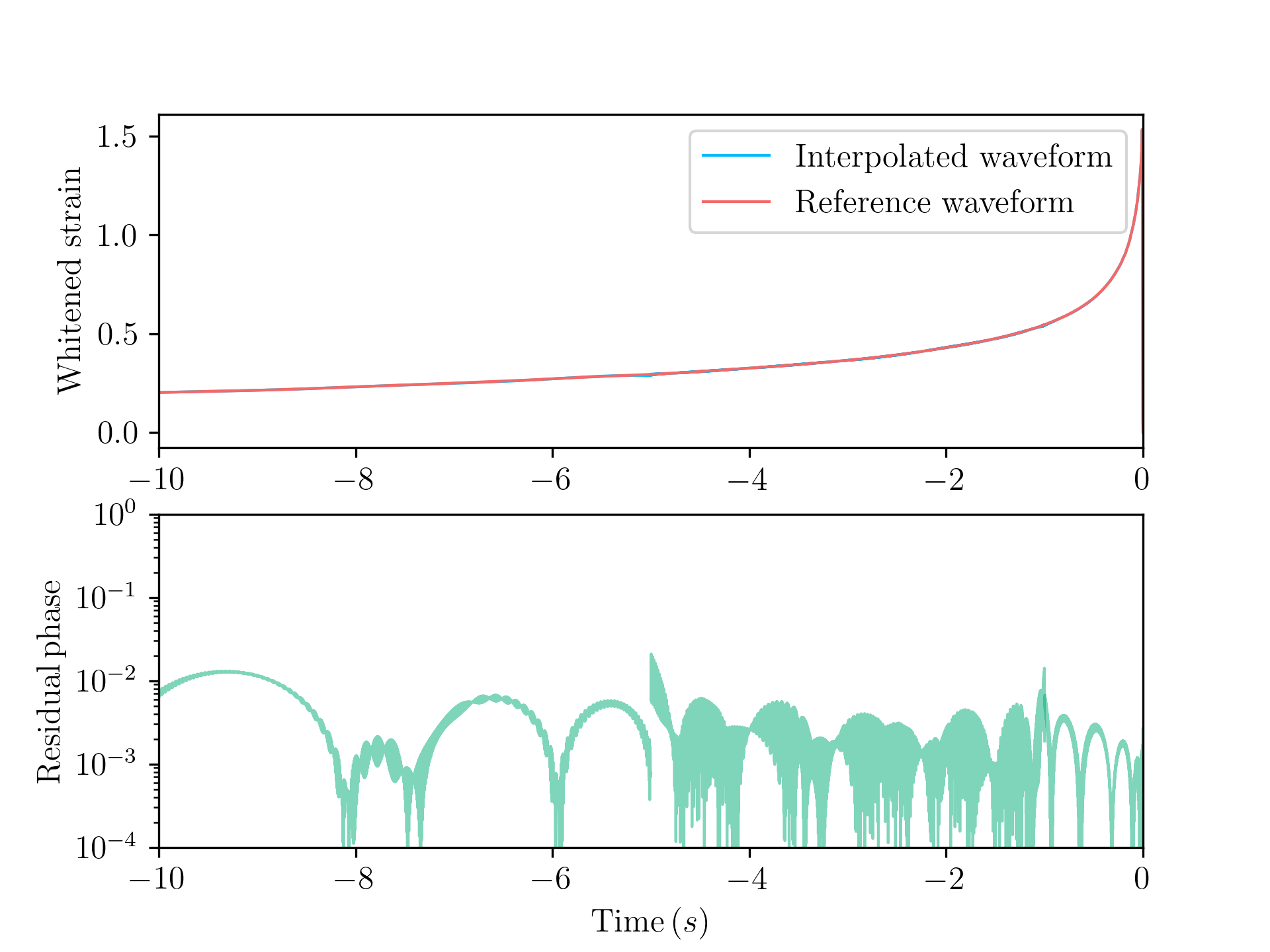}
	\label{fig:bns_residual}
}
\caption{Two representative interpolated waveforms and the associated residual
phase. The vertical dashed line in the left plot indicates the merger time for
the BBH. The pre-merger BBH phase agrees to approximately 1 part in
$10^1-10^2$, while the BNS phase agrees to 1 part in $10^2-10^4$. This shows that the interpolant provided here accurately reconstructs both the amplitude and phase of arbitrary signals.}
\end{figure*}

We train a neural network for each bank separately using the architecture
described in Section~\ref{sec:ml}. We find that both the BBH and BNS models are computed in
$\lesssim 10 \, \mathrm{minutes}$ on a GPU. The
performance of the resulting models are summarized in
Fig.~\ref{fig:mismatch}. We find that our
BBH (BNS) model achieves a median mismatch of $10^{-4}$ ($10^{-5}$). The
worst mismatch is $10^{-2}$ ($10^{-3}$), which is of the order of
waveform systematics~\cite{Bohe:2016gbl}. The worst reconstructed waveforms are
shown in Figure~\ref{fig:bbh_residual} and~\ref{fig:bns_residual} alongside
their phase residuals.

\subsection{Waveform and SNR interpolation timing study}

The frequency domain, aligned spin waveform models \texttt{TaylorF2} and
\texttt{SEOBNRv4\_ROM} can be generated in 1.1 ms and 2.5 ms, respectively, on
a central processing unit (CPU) using \texttt{LALSuite} waveform calls.  Recent
work~\cite{Edwards:2023sak} has examined accelerating waveform generation with
\texttt{JAX}~\cite{jax2018github}, producing an aligned-spin waveform
in 0.4 ms on a CPU (0.14 ms with \texttt{vmap} integration) and 0.02 ms on a
GPU.  Other works have also applied neural networks to aligned-spin BBH
waveforms, finding that they can be produced in 0.1 ms -- 5 ms on
GPUs~\cite{Chua:2018woh,Khan:2020fso,Schmidt:2020yuu}

We find that the neural network based interpolant described here has similar
performance to the above methods when utilizing accelerated linear
algebra~\cite{xla} and just-in-time compilation. We summarize our findings in
Table~\ref{table:profiling}.  We find that we can produce a single waveform in
0.5 ms on a CPU and 0.3 ms in a GPU. GPUs also offer the ability to batch
waveform generation. On an A100 GPU, we find that we can produce $10^4$
waveforms in $\sim 0.8$ ms, which corresponds to an effective speed of 1
waveform per 80 ns. The waveform generation time is roughly independent of
signal length since the size of the output layer size remains approximately
constant across banks. In all cases, we randomly sample from within the confines of the reference template bank.

\begin{table*}[]
\begin{tabular}{|c|c|c|c|}
\hline
& CPU & GTX 1080 & A100 \\
& Total time (Time per waveform) & Total time (Time per waveform) & Total time (Time per waveform) \\
\hline
Single BBH & 0.52 (0.52) & 0.40 (0.40) & 0.28 (0.28)  \\
\hline
Batched ($10^4$) BBHs & 28 (.0028)  & 3.6 (.00036) & 0.84 (.000084)  \\
\hline
Single BNS & 0.11 (0.11) & 0.41 (0.41) & 0.29 (0.29)  \\
\hline
Batched ($10^4$) BNSs & 27 (.0027)  & 3.5 (.00035) & 0.9 (.00009)  \\
\hline
\end{tabular}
\caption{Approximate times in milliseconds to generate our interpolated version of \texttt{TaylorF2}
waveforms for BNS (top) and \texttt{SEOBNRv4\_ROM} waveforms BBH (bottom)
systems on a consumer and industry grade GPU. Timing results are averaged over
$10^4$ independent calls for both single and batched waveforms. The evaluation
time is approximately constant across waveform length since the number of coefficients needed to reconstruct waveforms remains of the same order of magnitude.} \label{table:profiling}
\end{table*}

Recall that under our chosen framework, generating a GW via a weighted average
of SVD bases is equivalent to calculating the SNR timeseries using a weighted
average of SNR timeseries collected for each orthogonal basis found by the SVD
(see Equation~\ref{eq:snr}). The timings described in
Table~\ref{table:profiling} are thus also representative of how long it takes
to calculate an arbitrary SNR using the orthogonal SNRs output by the LLOID
algorithm in low-latency. For reference, we compare to the cost of performing this correlation
on a CPU with \texttt{Bilby}~\cite{Ashton:2018jfp}.
Table~\ref{table:bilby-profiling} shows the average time to compute $\langle h(t) | d(t)
\rangle$ in the frequency domain for a fiducial 16 s (64 s) BBH (BNS) waveforms at 5
different sample rates. Naively comparing our timing results to those of \texttt{Bilby}
indicate a $1-30$ time speed-up for a single SNR calculation and an
$\mathcal{O}(10^4-10^5)$ speed-up for calculations batched on a GPU. In each case, we 
compare the time to produce an SNR value when either Bilby or our
method is evaluated at a known point in the parameter space. This method is similar
to reduced-order-quadrature (ROQ)
techniques~\cite{Canizares:2014fya,Smith:2016qas,Morisaki:2020oqk}, which offer
similar accelerations for much broader signal manifolds than the ones
considered here, though these are not directly integrated into the detection
process.

\begin{table*}[]
\begin{tabular}{|c|c|c|c|}
\hline
& \multirow{3}{*}{\texttt{TaylorF2}} & \multirow{3}{*}{\texttt{SEOBNRv4\_ROM}} & \multirow{3}{*}{\texttt{IMRPhenomD}}\\
& & & \\
& (64s) & (16s) & (16s) \\
\hline
2048 Hz & 6.2 & 6.7 & 5.1 \\
1024 Hz & 4.0 & 5.1 & 3.1 \\
512 Hz & 2.5 & 3.1 & 1.7 \\
256 Hz & 2.0 & 3.2 & 1.9 \\
128 Hz & 1.5 & 3.6 & 0.6 \\
\hline
Speed-up (single) & 5 -- 21  & 13 -- 24 & - \\
Speed-up (batched) & $\mathcal{O}(10^4) $ & $\mathcal{O}(10^4) $ & - \\
\hline
\end{tabular}
\caption{Approximate time in milliseconds to generate a waveform and compute $\langle h | d \rangle$ in
the frequency domain with \texttt{Bilby}. Each row corresponds to waveforms generated at different
sample rates, while the column denotes a specific waveform family and signal duration.
We provide results for a variety of sample rates to reflect that our method
uses time-sliced and downsampled waveforms at intermediate steps. In all cases,
we fix the detector response and only profile waveform generation and the inner
product calculation. In our framework, waveform generation is equivalent to
reconstructing physical SNRs from SNRs measured by the SVD bases. A naive
comparison of our model's waveform
generation time in Table~\ref{table:profiling} to the
extreme sample rates shown here suggests that single waveform SNR calculations
can be accelerated by factors of a few to tens. When using batched generation,
the acceleration is more extreme; note that this assumes the \texttt{Bilby}
calls are still done in serial.}
\label{table:bilby-profiling}
\end{table*}

\subsection{Signal loss from PSD drift}

\begin{figure}
\includegraphics[width=\columnwidth]{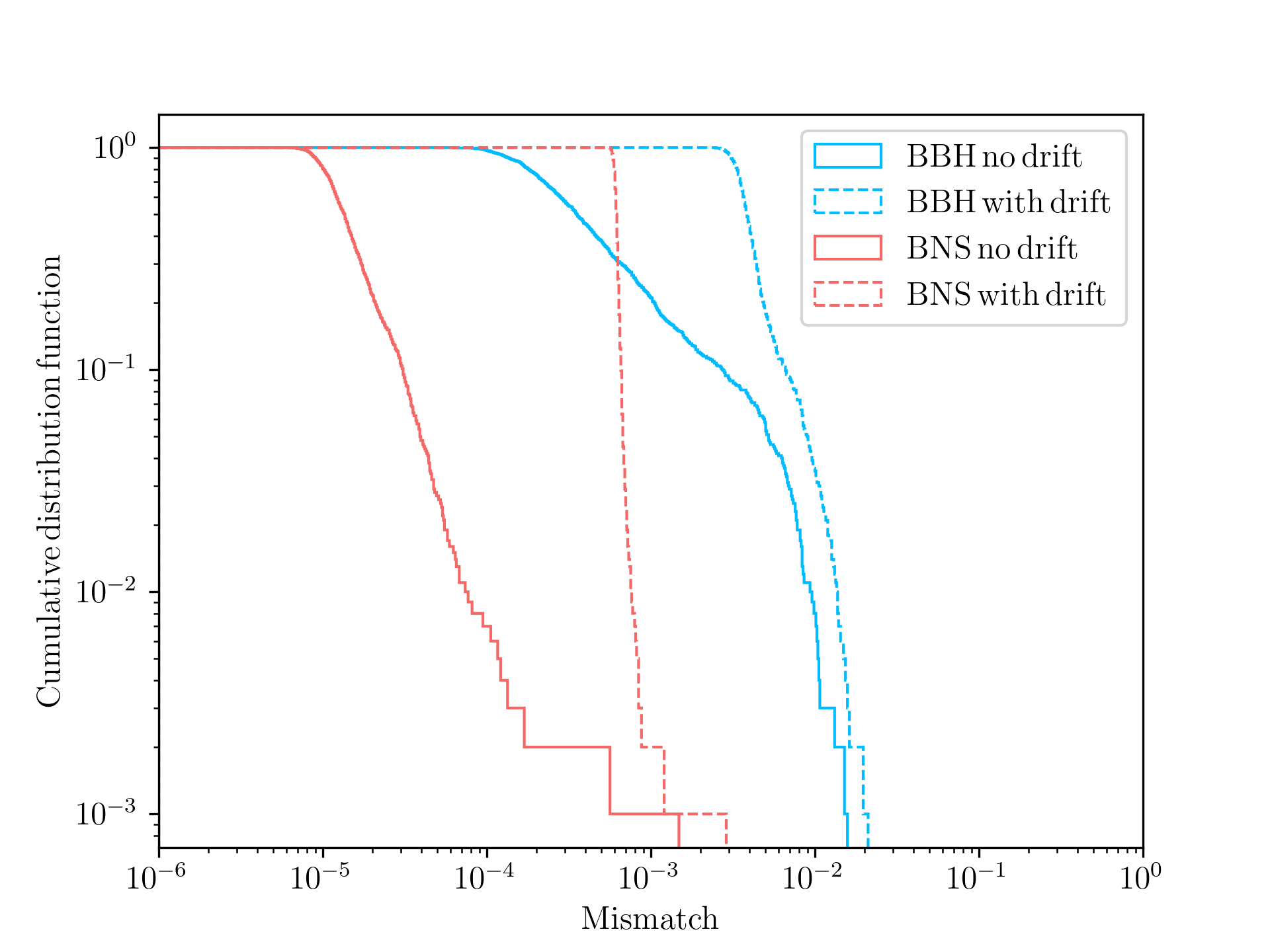} \caption{The mismatch
distribution for interpolated BNS waveforms for the PSD used in this work
(blue) and a more sensitive one (orange). While the bulk of the distribution
shifts towards higher mismatches, the interpolated waveforms remain accurate to
1 part in $10^2 (10^3)$ and the median mismatches remain of the same order of
magnitude as before. This is still of the order of expected disagreement
between different waveform families, suggesting that the models are still
appropriate for use.}
\label{fig:psd_drift}
\end{figure}

Since the basis vectors are constructed via SVD on the \emph{whitened} gravitational
waveforms, changes in the measured PSD fundamentally alter the bases
and can lead to signal loss. To minimize bias, the low-latency GstLAL analysis recomputes the
bases on a weekly cadence with updated PSDs. While in principle, the same
should be done for our method, we find it to be a small effect in practice. To
estimate the impact, we use the more optimistic, publicly available \texttt{aligoO4\_high.txt} PSD estimate for
O4\footnote{This PSD is also available at
\href{https://dcc.ligo.org/LIGO-T2000012/public} and represents LIGO detectors that are $12.5\%$ more sensitive than those described by \texttt{aligoO4\_low.txt}}. This is an extreme case of PSD
drift since the new PSD represents a detector that is $12.5\%$ more sensitive.
We use this as a proxy for PSD evolution over the course of the run.  With this
new PSD, we generate $1000$ sample whitened waveforms and compare to the
waveforms predicted by our neural network. We find that models trained on the less sensitive PSD still
reproduce the correct waveform to within 1 part in $10^2$ ($10^3$). These
results are shown in Figure~\ref{fig:psd_drift}. This is still within the range
of waveform systematics, though we do not rule out the possibility that
specific, frequency dependent PSD changes could lead to more substantial
differences. In the case of model breaking PSD changes, our network
can be asynchronously updated and retrained with the new spectrum. 
We leave a study of potential biases on parameter estimation to a
future work.

\section{Conclusion and future work}
\label{sec:conclusion}

In this work, we have presented a neural network based interpolant for GWs
connected to existing low-latency detection algorithms and data products. This
method allows us to accelerate the production of GWs and their associated SNR
time-series by up to a factor of $10^4$ for aligned spin binary systems while
retaining a faithfulness, on average, of 1 part in $10^4 (10^5)$ to the
original whitened BBH (BNS) waveforms. Although changes in the detector noise
properties reduce our model's fidelity, we find that it is still accurate to
within the level of waveform systematics.  The interpolant is constructed in a
highly parallel fashion to mimic how matched-filter based search pipelines
split the target parameter space to efficiently carry out low-latency analyses.
This leads to models that train in $\sim 10$ minutes. Modern searches cover
much broader parameter spaces than the two regions we consider here; we
estimate that it would take $\sim 1$ week to
train models in serial on a single GPU to cover the full parameter space
associated with matched-filter based analyses~\cite{Sakon:2022ibh}.

In low-latency, the GstLAL search pipeline functions as a collection of
microservices that loosely communicate with one another. Each microservice
controls the filtering (among other tasks) for small groups of sub-banks. The
microservices perform matched filtering in the reduced space described by the
SVD bases, but combine these outputs to provide a physical SNR associated with
the parameters of some template within the bank.  This work shows that the
intermediate SNRs collected by the basis vectors can be mapped to arbitrary
points in the sub-bank in near-real-time. If a new microservice is created to
ingest small snippets of intermediate SNRs from each filtering job, then the
lightweight model described here can be used to rapidly estimate the SNRs
at arbitrary points in the search space. This effectively amortizes the costly
likelihood calculation utilized by Bayesian inference based parameter
estimation pipelines and provides the SNR time-series across the search space
at arbitrary density.

This has the potential to both accelerate and improve
the accuracy of near-real-time parameter estimation based tasks. Low-latency
localization is performed using \textsc{bayestar}~\cite{Singer:2015ema}, which
fixes the masses and spins to the maximum search likelihood
estimate and performs Bayesian inference over the extrinsic parameters.
Though
accurate, it is well known that full parameter estimation produces
systematically more accurate skymaps, albeit at significantly greater
computational cost. SNR interpolation would allow algorithms such as
\textsc{bayestar} to marginalize over masses and spins to achieve a higher
accuracy. Alternatively, it could be used to maximize the SNR via
brute-force. After identifying candidates in low-latency, some detection
pipelines employ an SNR optimization task to scan the local parameter space for
slightly louder triggers. This interpolator could be used similarly, while
probing a much denser space than current algorithms. While this will likely
identify candidates with $\lesssim 1\%$ higher SNR, it will still lead to
systematically tighter localizations.

Source classification using Bayesian inference is presently done with
medium latency ($\mathcal{O}(\mathrm{minutes})$) algorithms that perform
full parameter estimation~\cite{Rose:2022axr,Morisaki:2023kuq}, but it is
possible that these could be accelerated even more with our interpolant.
These algorithms exploit the separability of aligned-spin gravitational
wave signals to efficiently evaluate the GW likelihood for arbitrary extrinsic
parameters. The method described here could similarly accelerate likelihood
evaluations at arbitrary intrinsic parameters. Future work will discuss the
connection of this technique to the \textsc{RapidPE}
algorithm~\cite{Rose:2022axr,Pankow:2015cra}, which utilizes
adaptive mesh refinement~\cite{1989JCoPh..82...64B}, in the pursuit of
$\mathcal{O}(\mathrm{seconds})$ parameter estimation.

Finally, since the mapping between bases and physical parameters is possible,
it is also intriguing to consider if the response from the reduced bases can be
used with likelihood-free methods~\cite{tejero-cantero2020sbi} to amortize
inference entirely. Some parameter estimation algorithms already utilize
similar techniques~\cite{Dax:2021tsq,Chatterjee:2024pbj}. We leave this study
for future work.

\section*{Acknowledgments}

The authors thank Jacob Golomb for Bilby assistance and useful comments.
LIGO was constructed by the California Institute of Technology and
Massachusetts Institute of Technology with funding from the National Science
Foundation and operates under cooperative agreement PHY-1764464.
This research has made use of data, software and/or web tools obtained from the
Gravitational Wave Open Science Center (https://www.gw-openscience.org), a
service of LIGO Laboratory, the LIGO Scientific Collaboration and the Virgo
Collaboration.  Virgo is funded by the French Centre National de Recherche
Scientifique (CNRS), the Italian Istituto Nazionale della Fisica Nucleare
(INFN) and the Dutch Nikhef, with contributions by Polish and Hungarian
institutes.
This material is based upon work supported by NSF's LIGO Laboratory which is a
major facility fully funded by the National Science Foundation.
The authors are grateful for computational resources provided by the LIGO
Laboratory and supported by NSF Grants PHY-0757058 and PHY-0823459.
This research has made use of data or software obtained from the Gravitational
Wave Open Science Center (gwosc.org), a service of LIGO Laboratory, the LIGO
Scientific Collaboration, the Virgo Collaboration, and KAGRA. 
This paper carries LIGO document number \dcc{2400308}.

\clearpage
\bibliography{references}

\end{document}